# High-Throughput Computational Studies in Catalysis and Materials Research, and their Impact on Rational Design

Mohammad Atif Faiz Afzal[1], Johannes Hachmann[2,3,4]

[1]Schrödinger Inc., Portland, Oregon 97204, USA

[2]Department of Chemical and Biological Engineering, University at Buffalo, The State University of New York, Buffalo, New York 14260, USA

[3]Computational and Data-Enabled Science and Engineering Graduate Program, University at Buffalo, The State University of New York, Buffalo, New York 14260, USA

[4]New York State Center of Excellence in Materials Informatics, Buffalo, New York 14203, USA

## Contents

## 1. Introduction

In the 21st century, many technological fields have become reliant on advancements in process automation. We have witnessed dramatic growth in both research and industries that have successfully implemented a high level of automation. In drug discovery, for example, it has alleviated an otherwise extremely complex and tedious process and has resulted in the development of several new drugs. Over the last decade, these automation techniques have been adapted in the chemical and materials community as well with the goal of exploring chemical space and pursuing the discovery and design of novel compounds for various applications. As every technology/device is connected to the performance of the materials that constitute it, the choice of materials is crucial. This is especially true considering the high investment costs associated with setting up mass production of a particular material. If the choice of a material is flawed, then the technology/device that builds on it may fail as well and result in dramatic losses. The impact of new materials on industrial and economic development has been stimulating tremendous research efforts by the materials community, and embracing automation as well as tools from computational and data science have led to an acceleration and streamlining of the discovery process. In particular virtual high-throughput screening (HTPS) is now becoming a mainstream technique to search for materials with properties that are tailored for specific applications. Its efficiency combined with the increasing availability of codes, both open-source and scalable commercial software, and large computational resources makes it a powerful and attractive tool in materials research.

HTPS is the process by which large numbers of compounds are characterized and assessed in an automated fashion (e.g., in the drug discovery context for activity as inhibitors or activators of a particular biological target, such as a cell-surface receptor or a metabolic enzyme). In addition to experimental screening studies, in which the enumeration of candidate compounds and their characterization is performed by experimental means, the field of virtual HTPS has seen tremendous growth. A number of large-scale *in silico* screening projects have been conducted in the materials field over the past few years. The key challenge in discovering new materials is that their behavior is governed by complicated structure-property and structure-activity relationships [1-3], and that chemical space is practically infinite [4-6]. Traditional research approaches alone are increasingly ill equipped to meet these challenges, in particular since advanced materials systems require more and more intricate property profiles [7-9]. Recent efforts have demonstrated that the combination of HTPS and modern data-science techniques allows us to pursue a rational design and inverse engineering paradigm that promises to mitigate many of the prevalent inefficiencies, shortcomings, and limitations of traditional approaches [10].

A typical computational materials discovery effort includes four key stages (see Figure 1): (i) model development, (ii) candidate library generation, (iii) high-throughput screening, followed by (iv) data mining and informatics analyses.

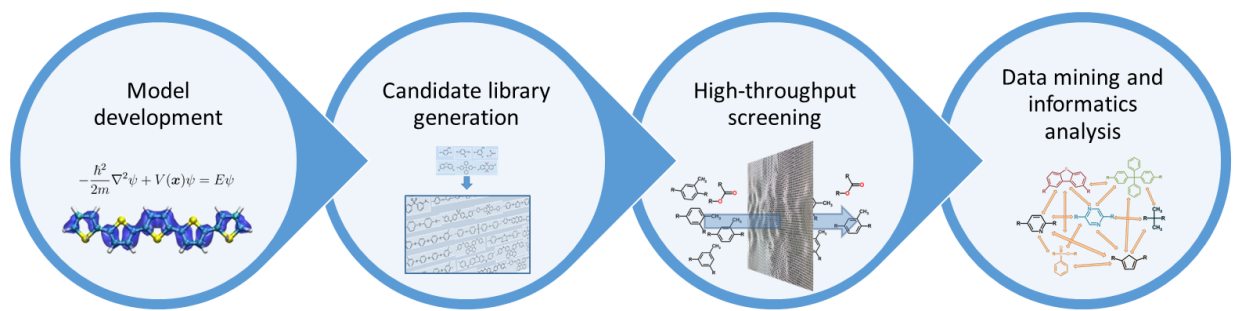

Figure 1: Key steps involved in a typical virtual high-throughput screening project.

**Model development:** The first step in the virtual HTPS process is the development of computational models or modeling protocols that allow us to predict the relevant properties of material classes of interest. Computational models can rapidly and efficiently characterize compounds, obtain key properties, and assess the performance potential of candidate compounds. These models can be developed using *first-principles* quantum chemical modeling, classical molecular mechanics and dynamics simulations, thermodynamic models, as well as cheminformatics-type quantitative structure-property relationship (QSPR) models. A key aspect in the model development is the benchmarking and validation of the proposed protocols to assess their predictive performance. Access to experimental data is not strictly necessarily, but highly beneficial in this context. Cost-accuracy analysis is another important issue. Once a sufficiently accurate and fast model for the assessment and scoring of candidate compounds is established, it can be employed to explore materials space to identify promising targets with desirable properties/performance for the targeted application.

**Candidate library generation:** A prerequisite for the high-throughput survey of materials space is access to suitable, large-scale screening libraries. These can be created based on a set of rules using a corresponding generator code for material candidate libraries. Other applications may require the enumeration of molecules or chemical reaction networks. A successful approach has to balance the ambition for a systematic and exhaustive enumeration of the combinatorial search space (which grows exponentially), with the need for a smart, responsive, and thus efficient scheme that focusses on the important regions of chemical space without wasting time on irrelevant candidates.

**High-throughput *in silico* screening:** For a very long time, the bottleneck in the overall HTPS process was the execution of large-scale computational studies. However, more recently, an unprecedented amount of computational resources as well as efficient codes have become available that renders *in silico* HTPS studies a viable proposition. HTPS codes provide an infrastructure that can automatize the setup and execution of thousands or even millions of calculations. They also have to be flexible enough to accommodate a variety of research fields, modeling and simulation engines, as well as hardware environments. A successful HTPS infrastructure implements automation at all available levels, including handling, parsing, and bookkeeping of the generated data, to make it as autonomous as possible.

**Data mining and informatics analyses:** HTPS studies result in vast amounts of data. In addition to the immediate information obtained from the screening studies (i.e., the identification of lead compounds that exhibit the desired property profiles), the generated data can be mined in its entirety using materials informatics and machine learning in order to facilitate a deeper understanding of the underlying structure-property relationships.

A general strategy to screening material candidates is based on a divide-and-conquer hierarchy, in which a given candidate library is filtered in a sequential process. This process employs a series of modeling protocols to evaluate different properties of interest and is typically sorted by computational cost or importance. As shown in Figure 2, the candidates are characterized, assessed, and screened at each level with respect to a different target property. The candidates are triaged and hopeless candidates discarded along the way, eventually resulting in a pool of candidates that fulfill all required characteristics within a predefined set of limits. Typically, the most computationally demanding properties are evaluated towards the end of this process where the number of candidates has already decreased significantly. Instead of filtering by different properties, a similar sequential process can be used to apply modeling protocols of increasing sophistication for the same property as the candidate pool is narrowed. This step-by-step process helps to zoom into the target region in materials space. More advanced and smart techniques, such as genetic algorithms and machine learning provide a path to further accelerating this process. We will discuss these techniques in detail in Section 4.

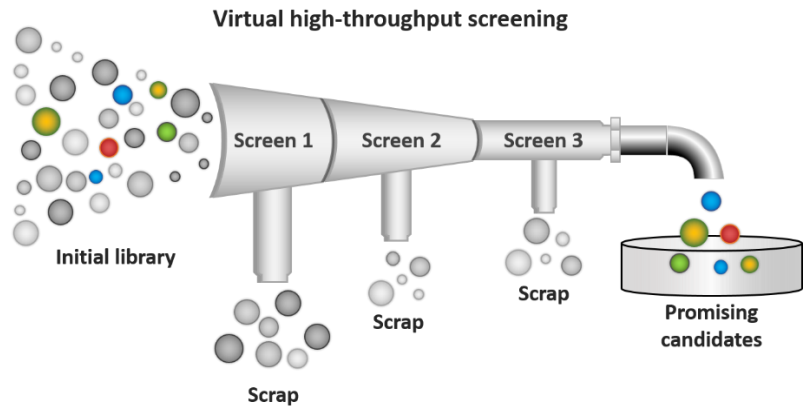

Computational HTPS has thus emerged as a promising approach to achieving the accelerated discovery of next-generation materials for various applications. Examples in which this paradigm was successfully implemented and utilized include the discovery of catalytic materials for various reactions [11], materials for energy storage [12], gas storage/separation [13], photovoltaics [14, 15], thermoelectrics [16], and OLEDs [17]. Some of the pioneering work in this field was performed by Nørskov, Persson, Ceder, Aspuru-Guzik, Snurr, Curtoralo, and others. In addition to the discovery of new materials, these virtual HTPS efforts provide a solid data foundation for rational design approaches as well as guidance for experimental collaborators.

In the following sections, we will review a selection of recent, high-profile HTPS projects for new materials and catalysts. In the case of catalysts, we focus on the HTPS studies for oxygen reduction reaction, oxygen evolution reaction, hydrogen evolution reaction, and carbon dioxide reduction reaction. Whereas, for other materials applications, we emphasis on the HTPS studies for photovoltaics, gas separation, high-refractive-index materials, and OLEDs.

## 2. Catalytic materials

One of the key challenges for the materials community is to design new catalysts and explore novel catalytic reactions [18, 19]. The area of catalysis has undergone considerable development, from discovery to automation, especially in high-throughput experimentation (HTE) [20, 21]. Even though these HTE approaches have been successful, they are limited by the exploration space. In these techniques, robots are capable of performing thousands of experiments; however, the outcome does not reflect the investments. This could be explained by the fact that robots are only capable of exploring limited grid search space, whereas the chemical space for designing catalysts is much larger [22]. Furthermore, the structure-property relationship for the catalysts are non-linear, which cannot be captured by grid search techniques [22]. Therefore, virtual screening techniques in the catalyst design space have received significant attention in the past decade.

The challenge in search of catalysts arises from the complexity involved in the chemical reaction pathways. Catalysis is a multi-dimensional problem where the process is affected by numerous variables, such as reactant's energies, configurations, and various transition states within each reaction step. The complexity of this process makes large-scale HTPS studies in catalyst design very challenging. To date, only medium scale throughput studies have been reported in the literature. The challenge lies in a clear understanding of the reaction mechanism and establishing chemical descriptors to evaluate the catalytic efficiency. Thus, a prerequisite to performing HTPS studies for catalysts is a thorough preliminary study on the reaction of interest [23]. Advances in density functional theory (DFT) and other computational methods evolved so many modeling techniques that can discover descriptors of catalytic activity and other structure-property relationships. These can be extended over a library of candidate catalysts for their evaluation, thus leading to a form of high-throughput computational screening. Several research groups have applied this approach to model chemical reactions, e.g., hydrogen evolution reaction (HER) and oxygen evolution reaction (OER) over metal surfaces [11, 24-30]. The aim of these studies was to map the catalytic activity to a single descriptor, for example, the adsorption energies of key reaction intermediates to the catalyst surface. The agreement between these modeling studies and the experimental data was striking, as was the mapping, which confirmed that the most active catalysts had favorable values of the descriptors. It is, therefore, necessary to find useful structure-property relations that will serve as sufficient descriptors of catalytic activity for the discovery of new catalysts.

Identifying descriptors of catalytic activity involves three major steps. The first step is discerning the elementary steps involved in the reaction mechanism. The second step includes evaluation of the energetics of the individual steps involved in the reaction mechanism. For example, in heterogeneous catalysis, this would include the dissociation and adsorption energies and reaction barriers for all the transition metal surfaces *via* ab initio calculations. This is followed by the identification of active sites, the sites with the lowest energy barrier. The third step is identifying scaling correlations between adsorption energies and corresponding activation energies of the transition states [31, 32]. These scaling relationships serve as descriptors and can be used to quantify catalytic activity without having to evaluate all thermodynamic parameters involved in the complete mechanism. Once the reaction steps are understood and the descriptors are identified, the next step involves evaluating the activities for a library of promising candidates.

The activity of a catalyst is based on its electronic structure, and therefore the catalytic performance can be improved by tailoring their chemical structure. By taking the HTPS approach, it is possible to tailor the structure of catalyst in a large-scale fashion. Several HTPS efforts have been undertaken in the last decade in search of new high-efficient, low-cost, and environmentally friendly catalytic systems for various applications.

## a. Oxygen reduction reaction (ORR)

The oxygen reduction reaction (ORR) plays an important role in energy conversion, biological respiration and material dissolution [33]. Example processes include lithium-air batteries and fuel cells [34, 35], as well as Polymer Electrolyte Membrane Fuel Cells (PEMFCs) [36]. PEMFCs are characterized by high-energy conversion rate, power density, and are environmentally friendly. Performance of PEMFCs is primarily dependent on the catalyst, a critical component in the membrane electrode assembly. It has been shown that Pt catalysts are the most effective for ORR [37]. However, the high raw material cost currently prevents the commercialization of this technology. Therefore, the development of active and stable low-cost materials to replace Pt catalysts has become the main challenge in creating viable PEMFCs. Initial attempts to replace Pt, evaluated the use of Au, Ag, and Pd-based alloys [38, 39], 2-dimensional (2D) materials such as graphene sheets and dichalcogenides [23, 40-44], and nanoscale alloy particles [45, 46].

There are primarily two pathways for ORR: the two-electron pathways to form hydrogen peroxide and the four-electron pathway to form water [47]. The latter pathway is the most commonly observed and has been extensively studied in surface electrocatalysis. This pathway consists of four intermediate steps:

1) reduction of $O_2$ to form HOO* ($O_2$ +*+$H^+$+$e^-$ $\rightarrow$ HOO*),

2) reduction of HOO* to form O* and release of a water molecule ( HOO* + $H^+$ + $e^-$ $\rightarrow$ $H_2O$ + O*),

3) reduction of O* to form HO* (O* + $H^+$ + $e^-$ $\rightarrow$ HO*),

4) further reduction of HO* to release a second water molecule (HO* +$H^+$+ $e^-$ $\rightarrow$ $H_2O$ + *).

Among these four steps, the potential determining steps are the first and fourth. There exists a scaling relationship between the free energy of HO* and HOO*, therefore, either of these reactions could be considered to define an ORR descriptor. In the past, different screening studies have used different descriptors to quantify the ORR catalytic activity. For example, one study implemented the free energy of adsorption of oxygen as a descriptor [28], whereas another used the free energy of HO* [48].

Thorough preliminary studies on ORR reaction and the discovery of descriptors allowed for several screening efforts for the ORR catalyst design. It should be noted that, due to the complexity of the ORR reactions, the scale of many of the screening efforts is low to mid-throughput.

One of the first computational screening studies for the identification of heterogeneous metal alloy catalyst for ORR was performed by Nørskov and co-workers [28]. They performed DFT calculations on 750 binary alloys of transition metals to compute their activity and identified several promising candidates. However, these promising candidates were observed to be thermodynamically unstable, using rigorous, potential-dependent stability tests. Even though their work did not result in viable candidates, they found an efficient screening formalism for evaluating the catalytic activity and further established stability criteria. Their screening approach can be applied to several other systems in a systematic manner to identify candidates that are both highly active and stable for practical fuel cell applications. In a recent study, similar approach of activity and stability based screening was performed on Pd-based catalysts [49]. They reported that Pd-V, Pd-Fe, Pd-Zn, Pd-Nb, and Pd-Ta alloys have high stability and improved ORR activity.

There has been growing interest in the use of carbon-based materials (e.g., graphene sheets and carbon nanotubes [42, 50, 51]) as catalysts for ORR. Notably, such materials are being extensively used as interfaces for creating metal-embedded semiconductor composites. These materials possess a large electron-storage capacity, good electron conductivity, stability, and chemical and mechanical strength. Favorable charge transfer properties accelerates charge carrier separation at the semiconductor and transfer to the catalytic reaction sites. Graphene sheets doped with heteroatoms, such as N, B, S, P atoms, have demonstrated enhanced catalytic activity [50, 52]. Additionally, these doped graphene sheets are highly stable and inexpensive compared to the Pt-based catalysts [50]. The reason for enhanced catalytic activity of these 2D materials is the spin density and charge density redistribution around the heteroatoms, which arises due to the electronegativity differences between heteroatoms and carbon [44]. The redistribution of spin density and charge density allows for the chemisorption of $O_2$ molecules, thus aiding in the breakage of the O-O bond.

Doping of graphene with other elements result in the generation of more charge carriers in the system which helps carry out the reactions and further increase the activity. Recently, Jiao *et al.* studied a series of graphenes doped with non-metal elements and evaluated their performance based on four descriptors: exchange current density, on-set potential, reaction pathway selectivity and kinetic current density [43].. Based on the DFT calculations, they derived a volcano plot, similar to the plots observed in the case of metal catalysts, between the ORR performance and the free energy of OOH* adsorption. The screening studies suggested that graphene-based metal-free catalysts are highly promising for ORR and have the capability to surpass the catalytic activity of Pt catalysts. In another recent study, van der Waals corrected DFT was used to screen metal decorated graphenes for improving the ORR activity [41]. The free energy of monoatomic oxygen was used to rank the activity of the catalysts. In their work too, they show a volcano plot trend, which indicates Au dominant Pd intermetallics have the most catalytic efficiency, i.e. more than Pt-Pd intermetallics. In addition to the chemical composition, they found that the size of nanoparticles also plays a role – an order of magnitude change in particle size resulted in a significant improvement in the catalytic activity.

Another category of 2D layered materials, transition metal dichalcogenides (TMDs), are also inexpensive electrocatalysts with promising physical and chemical properties [44]. Wang *et al.* performed computational screening studies on TMDs, particularly on 2D $MoS_2$ monolayers, to identify the best catalytic activity for the ORR [44]. They systematically evaluated the catalytic activity of $MoS_2$ monolayers doped with a series of transition metals (TM, V, Cr, Mn, Fe, Co, Ni, Cu, Nb, Ru, Rh, Pd, Ag, Ta, W, Re, Os, Ir, Pt, and Au). The results show a linear relationship between the calculated $\Delta G$ of OOH* and OH* on various $MoS_2$-based catalysts (see Figure 3). The DFT results further demonstrate that the transition metal atoms strongly interact with the S-vacancy, which modifies the electronic and magnetic properties of $MoS_2$ monolayer surface. These small-scale DFT-based screening studies revealed that $MoS_2$ monolayer embedded with Cu, which shows optimal binding strength with the ORR intermediates, has the best catalytic activity due to its minimum overpotential of 0.63 eV.

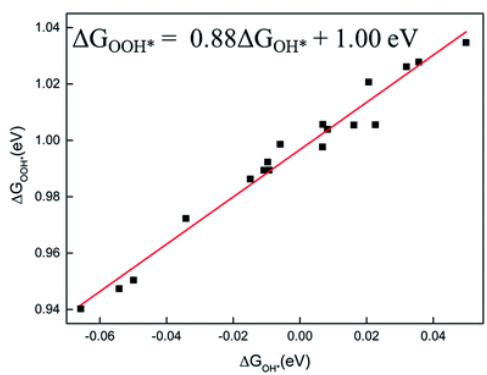

Nano-scale alloy catalysts have also been explored for ORR catalytic activity [51]. These systems are attractive since they require low Pt loading and are shown to have efficient catalytic activity due to enhanced active surface sites [53]. Nano-scale catalysts consist of a core-shell like structure with the core consisting of transition metals and the shell consisting of Pt. It was shown that ternary alloy core-shell catalysts, such as PtNiCu and PtCuCo, have long-term stability and catalytic activity in comparison to binary alloys [54]. Furthermore, these ternary alloys have features suitable for commercialization. An effective approach to design ternary alloys with superior catalytic properties is to computationally screen catalysts with varying alloying metals in the core. This will provide a better understating of interactions between different shells, thus allowing rational design of highly stable and durable catalysts. In a recent study, Noh *et al.* performed DFT calculations on 158 different nano-scale catalysts with varying binary alloy core, using Fe, Ni, Cr, Cu, and Pt on the surface (see Figure 4a) [51]. Their results indicated that $Pt_{skin}Cu_{0.76}Ni_{0.24}$ nanoparticle of 3nm size had better electrochemical stability than pure Pt catalyst (see Figure 4b). It was proposed that the enhanced catalytic activity is associated with the compressive strain on Pt surface and the increased electrochemical stability of the catalyst is due to the interactions among the nanoparticle shells. The screening approach in their study can aid in the design of high-performance catalysts for PEMFCs and can also be applied for other kinds of catalytic materials in a similar electrochemical environment.

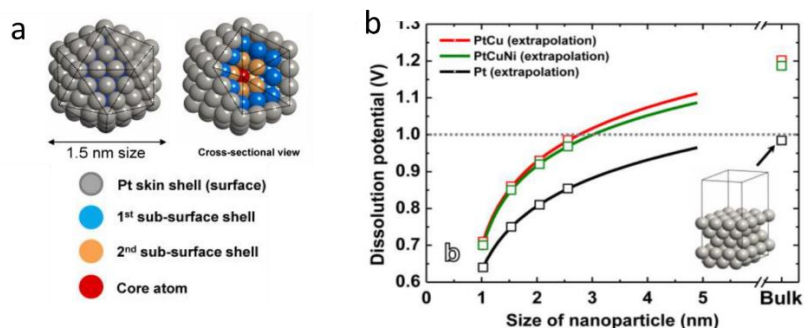

**Figure 4: (a) Illustration of Pt skin nanoparticles. The nanoparticles consist of Pt skin, 1st layer, 2nd layer, 3rd layer, and core atom. (b) Dissolution potentials as a function of the size of nanoparticle of Pt, PtCu, and PtCuNi and their extrapolations into larger particles. Figure and caption reproduced from Ref. [51].**

## b. Oxygen evolution reaction (OER)

The oxygen evolution reaction (OER) is an important reaction in the process of water splitting. The splitting of water molecules *via* photocatalytic reaction gained momentum as a complementary approach to photovoltaic, photothermal, and photosynthetic production of electricity, heat, and biomass, respectively. The direct production of fuel (i.e., chemical energy) avoids the storage and transport issue of electrical or thermal energy, and can be used in fuel cells or exhaust-free combustion replacing non-renewable energy carriers.

In principle, the process of splitting water can be divided into two reactions, namely OER (water oxidation) reaction at the anode and HER (water reduction) at the cathode. Of these reactions, OER is thermodynamically unfavorable, making this reaction highly critical from kinetics point of view. Formation of the oxygen-oxygen bond involves the removal of four protons and four electrons from the water molecule, and the whole process comprises of four intermediate steps. The process is a reverse reaction of the aforementioned ORR reaction. In the first step, water is oxidized on one of the active sites releasing one proton and one electron, resulting in the formation of HO* intermediate on the surface. This intermediate is subsequently oxidized to form O*. A second water molecule then reacts with O* to form superoxide intermediate, HOO*. This new intermediate is further oxidized to form $O_2$. Similar to the ORR reaction, there exists a linear relationship between the free energy levels of HO* and HOO* [55]. This allows for the selection of a single descriptor for OER catalytic activity. This descriptor, overpotential, is used as the ranking criteria in the screening of new catalytic materials for OER.

The idea to chemically capture solar energy is inspired by natural photosynthesis in green plants and cyanobacteria [56, 57]. In the natural photosynthetic process, OER reaction is catalyzed by $Mn_4Ca$ clusters. Therefore, it was earlier believed that the catalysts with multinuclear metal centers were required to catalyze OER. As a result, a large number of multi-metal complexes such as tetra/di manganese [58], tetra-cobalt [59], and tetra/di ruthenium [60], were extensively studied. However, a computational screening method for the discovery of catalysts with two or more different metal centers is not possible *via* a single descriptor approach. In the case of homobimetallic species, a Sabatier analysis allows a single descriptor for predicting the catalytic activities and realizing trends expected for the first-row transition metal elements. However, in the case of heterometallic systems where two different metals act as active centers, the Sabatier analysis cannot be applied. Based on the DFT calculations of various heterobimetallic catalysts, it was confirmed that a single chemical descriptor is not sufficient to describe overpotential trends and the mixed-metal overpotentials cannot be predicted based on the pure-metal redox potentials [61]. The inability to find a single descriptor makes it difficult to perform HTPS studies on heterometallic catalysts. Therefore, in such cases, we have to expand beyond the single descriptor analysis iand further develop accurate models to compute the redox potentials of these complex heterometallic catalytic reactions.

Interestingly, it was later shown that it is possible to catalyze OER using mononuclear catalysts [62]. These catalysts have similar catalytic efficiency as the multi-metal complexes and are significantly easier to design, synthesize and characterize. Additionally, single metal sites are easy to evaluate computationally, making them interesting targets for HTPS. Furthermore, the relationship between the electronic structures and the molecular geometries can be systematically studied and the effect of ligand on the efficiency of catalysts can be easily tailored. These remarkable attributes of mononuclear catalysts motivated researchers to perform HTPS studies.

Manganese-based compounds have been extensively studied for OER due to their relevance in photosystem II, which is the active center of the photosynthesis process. Photosystem II consists of $Mn_4O_4Ca$ cubane-like structures that are enveloped by large protein chains. Additionally, Mn is non-toxic, abundant, and it has the ability to form mixed oxides due to its multiple valences. Mn provides optimal binding energies for various OER intermediate reactions, thus lowering the reaction overpotential. In a recent study, $\alpha$-$MnO_2$ doped with several transition metals were computationally screened [63]. The preferred valence at each site in that study was enforced by addition/removal of hydrogen and hydroxyl groups. In most cases, lower overpotentials were observed on closely packed (110) surface. Furthermore, they identified three different active sites (cus, bridge, and bulk) and demonstrated that the dopants prefer the surface over the bulk sites. $\alpha$-$MnO_2$ doped with Pd exhibited the best catalytic efficiency for OER.

Nickel hydroxide based materials have also shown to have good catalytic activity for OER [64-66]. Additionally, the activity of Ni hydroxide was shown to improve when combined with other transition metals [67]. This inspired a systematic study to understand the effect of various transition metals on the catalytic activity of Ni-based materials for OER [68]. This low-throughput screening study defined simple guidelines for the rational design of Ni-based catalysts for OER. It was shown that Cr, Mn, and Fe improve the catalytic activity of the Ni-based double hydroxides, whereas Ni hydroxides with Co, Cu, and Zn have poor catalytic activity. Ni-doped with Mn, Fe, Co, Cu, and Zn resulted in slight increase in the OER overpotential of Ni sites, while Ni-doped with Cr showed a decrease in the OER overpotential (see Figure 5). Among Fe, Mn, and Cr, the active sites in NiFeOOH and NiMnOOH were Fe and Mn, respectively, and Ni was the active site in NiCrOOH. In addition to theoretical studies, they synthesized these catalysts and confirmed their catalytic activities experimentally.

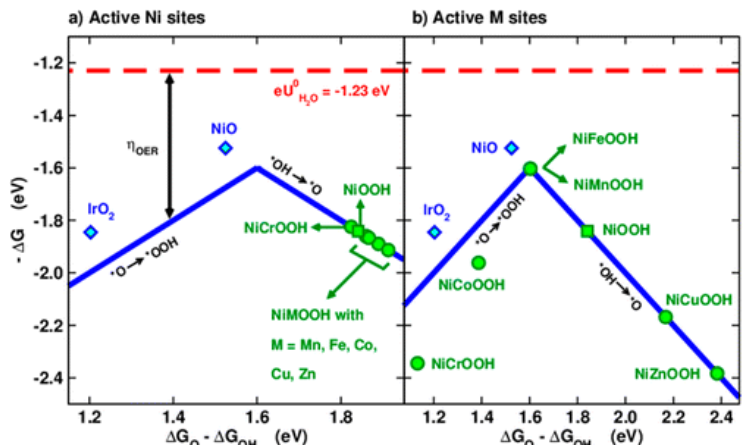

Most of the earlier research in catalysis has focused on metals or metallic oxides and their derivatives [69]. However, recent attention has shifted towards the vast chemical space of organic electronic compounds and metal-organic complexes such as corroles, porphyrins, aromatic diimides, and bisimides etc. (see Figure 6) [70]. These compounds are particularly interesting because they can be manipulated on the molecular level to tailor their properties. These compounds have also shown the propensity to catalyze OER reactions, albeit with different functional groups. This indicates a potential trend where we might hypothesize that corrole and porphyrins with electron donating groups favor the reduction reactions and systems with electron withdrawing groups favor the oxidation reactions. In addition to metal-organic complexes, several 2D materials have also been studied for OER [71-75]. The 2D network formed by benzene rings give these materials similar properties to corroles and porphyrins, which constitute the prevalent cofactors used in natural photosynthesis. Thus, favorable electronic properties are combined with superior surface morphology to give improved catalytic activity. As mentioned before, they have shown huge improvements over expensive Pt catalysts in the case of ORR in fuel cell systems [76]. The enhancement in OER activity is also promising when such catalysts are used [77-79].

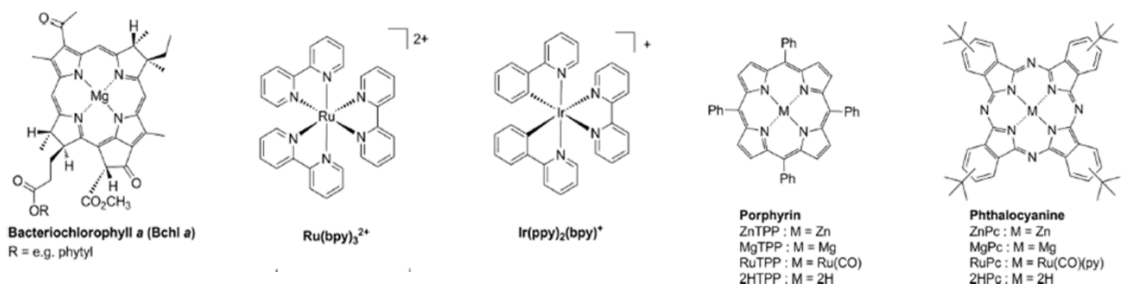

A distinctive feature of these materials is the ability to tailor their properties by enabling minute changes to the structure using ion doping/intercalation, mechanical straining, and edge/defect engineering [80]. Tailoring such doped graphene-like structures, along with the porphyrin and corrole based macrocyclic compounds, could potentially lead to a large number of potential candidates. Recently, some of these 2D materials have been explored for OER potential, but to the best of our knowledge no large-scale screening studies have been implemenetd yet. Therefore, based on the establishment of a single descriptor, there is a huge potential for performing HTPS studies to explore new 2D materials for efficient catalysis in OER.

### c. Hydrogen evolution reaction (HER)

Hydrogen evolution reaction (HER) is an important electrochemical reaction for various applications including hydrogen fuel cells, electrodeposition, corrosion of metals in acids, and energy storage by the generation of $H_2$ from water splitting. HER had been thoroughly investigated in the past and well-defined atomic-scale descriptors have also been identified. Following the Sabatier principle, a volcano shape is observed when the catalytic activity of a material for HER is plotted against the hydrogen-metal bond strength (see Figure 7) [27]. This plot shows that the free energy of adsorption of hydrogen on the surface is a good descriptor to quantify the catalytic efficiency. High binding energy results in strong surface adherence, whereas low binding energy results in less hydrogen availability – both of the cases leading to poor catalytic efficiency. Thus, optimal hydrogen binding energy is required to achieve high catalytic efficiency.

Greely and co-workers pioneered the above-mentioned descriptor to perform HTPS for HER catalysts, which included the screening of 736 distinct binary transition-metal surface alloys [81]. The catalysts were ranked based on the free energy of adsorption, i.e., the closer the free energy to zero, the better the catalytic efficiency, as demonstrated in Figure 7. A comparison of the activity of 256 binary surface alloys, as schematically plotted in Figure 8a, shows that numerous binary alloys have high predicted HER catalytic activity. In addition to the activity, they computed the stability of the alloys based on four tests: estimation of the free-energy change associated with surface segregation events, island formation and surface de-alloying, oxygen adsorption, and the likelihood of dissolution of the alloy in acidic environments. Based on these tests, they identified several candidates that are both stable and active (see Figure 8b). Some of the top surface alloys identified are BiPt, PtRu, AsPt, SbPt, BiRh, RhRe, PtRe, AsRu, IrRu, RhRu, IrRe and PtRh. This work demonstrated that stability considerations are essential for finding candidate catalysts that are synthetically feasible. Since this work by Greely, there have been several high-throughput efforts that apply a similar approach to identify HER catalysts [75, 82-84].

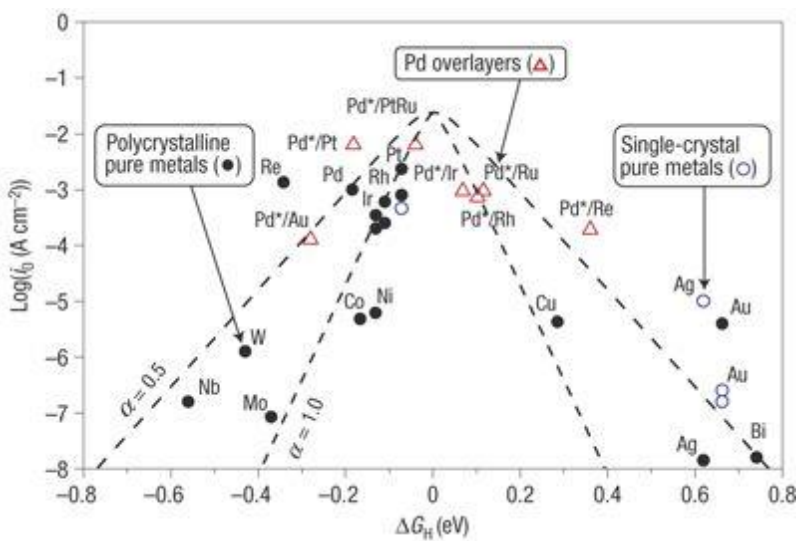

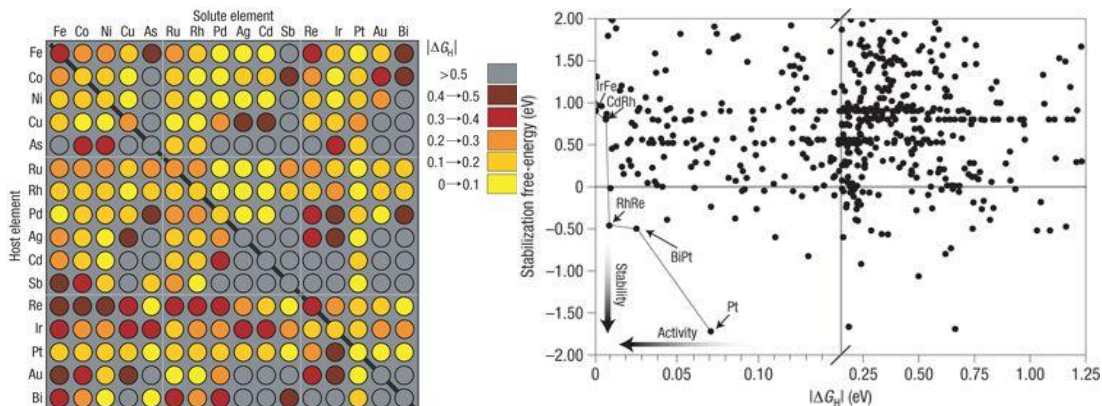

Similar to the ORR and OER reactions, 2D materials have also shown a promise for HER catalysis [85, 86]. A new type of 2D materials, MXenes, which comprises of carbides and nitrides of transition metals, have shown huge potential in HER catalysis. The ability to control the thickness of these 2D systems enabled exploration of a large design space. Pandey *et al.* performed screening studies on the HER catalytic activity and the stability of MXenes of the type $M_2X$, $M_3X_2$, and $M_4X_3$, where M is a transition metal and X is either N or C [87]. The catalytic activity was evaluated using the same descriptor as mentioned in the previous study, i.e., using the free energy of hydrogen adsorption (see Figure 9), and the stability is evaluated by calculating the heat of formation of MXenes. Their screening results confirmed that the thickness of MXene is an important parameter, which could be tailored to maximize the catalytic efficiency. Furthermore, stability studies revealed the importance of functionalizing agents in synthesizing thermodynamically stable MXenes. Several MXenes with both high activity and stability were identified using these screening studies. MXenes terminated with oxygen were shown to have the best catalytic activity, which was further confirmed by a screening study performed by Ling and co-workers [82]. Tailoring the surface functional groups and the transitional metals of MXenes could yield enhanced catalytic activity. Using this approach, a recent screening study found that $Mo_2C$ based MXenes exhibits far better catalytic activity than Ti-based MXenes, which was further confirmed *via* experiments [88].

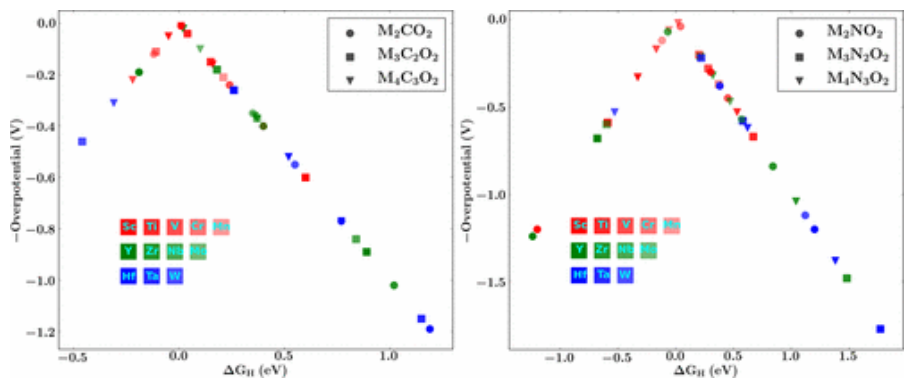

Inexpensive pyrites, which are composed of the first-row transition metals and dichalcogenide ligands ($MX_2$, where M = Fe, Co, or Ni and X = S or Se), have also showed good catalytic activity for HER [89]. Even for $MS_2$ and $MSe_2$ type pyrites, the free energy of hydrogen adsorption is a good descriptor [89]. Screening studies demonstrated that the catalytic activity in pyrites is based on the location of d-band center energy of the transition metal, the orbital energy split by the ligand field, and the electron pairing penalty in the d-orbital at the same energy level. Furthermore, the stability analyses confirmed that the transition metals in pyrites are thermodynamically stable against electrochemical degradation, thus, making them highly promising for HER catalysis. All the above-mentioned screening efforts exploring pyrites and MXenes for HER catalysis have paved a way for such materials in other clean energy reactions as well.

The wide variety of systems subjected to prior research for ORR, OER, and HER and the broad description of catalysis available through the use of theoretical descriptors opens up a world of possibilities of further investigation in this field. Using these established reaction mechanisms and scaling relations, one can explore chemical space with efficient high-throughput algorithms and locate more candidates that are interesting.

### d. Carbon dioxide reduction reaction

One of the critical problems affecting climate change is the increasing $CO_2$ levels in the atmosphere. To mitigate the issue of anthropogenic climate change, the emerging technology of electrocatalytic reduction of $CO_2$ has shown immense potential. However, efficient catalysts are required to make this technology feasible. A large number of metal catalysts have been investigated as catalytic materials for the reduction of $CO_2$. So far, the catalytic activity of these metal-based catalysts is inefficient for practical applications, and therefore, numerous non-metal based catalysts have also been investigated. The reduction of $CO_2$ to Co is a two-electron step which includes the formation of two intermediates, COOH* and CO*. A strong positive correlation exists between the adsorption energies of these two intermediates, which makes it difficult to modulate individual binding energies. However, modification of the surface to include a covalent character can break this scaling relation, stabilize COOH*, and in turn lower the reduction overpotential. Layered materials, such as graphene/graphite and dichalcogenides, perform better than conventional metallic catalysts as they are able to conduct electrons despite the covalent bond network. A recent DFT based screening study of 61 2D covalent metals showed that the scaling relation can be entirely broken (see Figure 10) [90]. The screening results demonstrated that $IrTe_2$, $RhTe_2$, PFeLi, and $TiS_2$ are better catalysts than Au for CO production, whereas LiFeAs and $ScS_2$ show better catalytic activity than Cu for $CH_4$ production. Other recent screening studies for electrocatalytic reduction of $CO_2$ include metal-based catalysts, bimetallic catalysts, and zeolites [91-94].

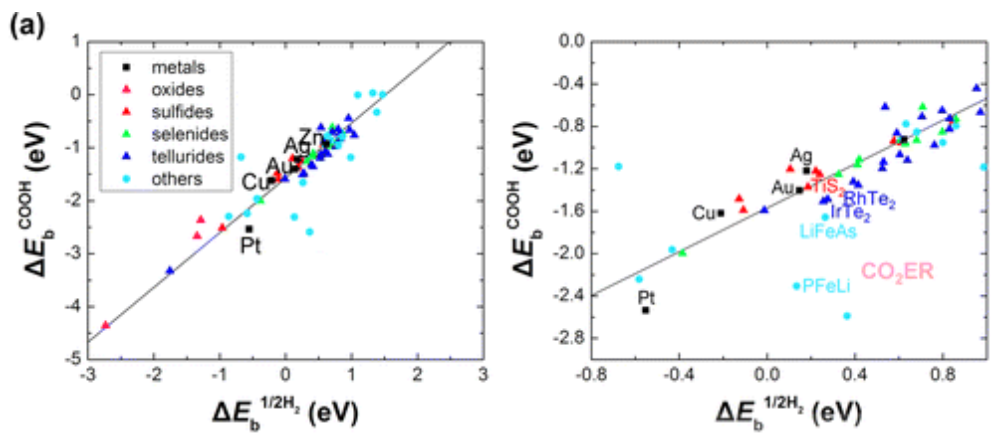

Figure 10: a) Relationship between the binding energies of COOH and H on the catalyst; b) Enlarged plot to emphasize the deviations from the linear relationship. Figure reproduced from Ref. [90].

Other notable reactions where HTPS has been applied successfully are ammonia production [95, 96], methane activation [97, 98], desulfurization [99], and acetylene hydrogenation [100]. In all these screening studies for catalysts, the level of screening is limited to medium-scale due to the complexity and high cost of the computational methods. However, computational resources are becoming cheaper and faster. Additionally, new developments in computational methods to study reactions allow for more accurate prediction of catalytic activity. These developments in computational availability and method accuracy could enable the promising avenue of performing large-scale HTPS for the discovery and development of efficient catalysts.

## 3. Other materials

### a. Solar materials

The increased use of photovoltaics has stimulated tremendous research in the materials community in the last decade due to their promise in dramatically reducing electricity generation cost [101]. Many countries are using photovoltaics technology as an electricity source to reduce their carbon footprint. However, it is still an expensive process in comparison to the conventional energy sources. Photovoltaics are still far away from reaching the current energy demand. To make this technology mainstream, we have to improve the efficiency, cost of synthesis, and the durability of the new photovoltaic materials [102]. To that end, recent research initiatives have focused on new breakthroughs which go much beyond the existing Si technology [103]. Rather than focusing on incremental improvements in power conversion, these novel concepts are targeting new breakthroughs by designing novel materials with power conversion efficiencies (PCE) that go beyond the Shockley–Queisser limit, have improved recyclability, and resistance to degradation under extreme conditions. The key challenge in designing such materials is understanding the complex relationship between PCE and the structure of both the active materials as well as the supportive materials in photovoltaic devices. HTPS approaches have shown to uncover these relationships and aid in the discovery of novel materials for photovoltaics [15, 104-109]. The two primary technologies where HTPS is being applied are the solution-processed technologies, such as organometal halide perovskites [106, 107] and organic photovoltaics [14, 15, 110, 111]. These technologies are highly promising due to their low cost of manufacturing and good power conversion efficiency.

Carbon-based materials have attracted significant attention as an alternative to conventional Si-based technology due to their low-cost, easy processability, flexibility, and lightweight. However, there are still several issues that are limiting their use in solar cells, like their relatively low efficiency and limited lifetime [112]. For such organic materials to be accessible in practical photovoltaic devices, their efficiency needs to be increased by 10–15% with a lifetime of greater than 10 years. The Harvard Clean Energy Project (CEP) was established to search for such organic materials by combining conventional modeling with strategies from modern drug discovery [14, 15]. This includes a systematic HTPS of millions of organic candidates for donor molecules at DFT level, and further applies techniques from cheminformatics and data mining to uncover structure-property relationships. The ranking of the candidates is performed by employing the Scharber model [113], which is a specialized version of the Shockley–Queisser model for organic photovoltaics [114]. The inputs for this model are the highest occupied molecular orbital (HOMO) and the lowest unoccupied molecular orbital (LUMO). Even though the Scharber model is too simple to account for the complex underlying physics, it provides a metric to identify promising candidates from the initial screening of a large candidate library. Within the Scharber model with a standard phenyl-C61-butyric acid methyl ester acceptor, the theoretical limit was calculated as 11.1%. The optimum HOMO and LUMO parameters required for this limit are -5.41 eV and -4.00 eV, respectively. HOMO and LUMO values of 2.3 million candidates screened in CEP project are shown in Figure 11. The range of HOMO and LUMO values for this library is quite broad, as apparent in Figure 11b. However, very few candidates are confined in the parameter space for high-performance materials (marked by the white circle). Among the 2.3 million compounds, about 1000 candidates exhibit a PCE of 11% and higher, while the majority of the candidates demonstrate a PCE of less than 4% (see Figure 11c). Using cheminformatics and data mining techniques, building blocks such as thiadiazoles and silaindenes were shown to have ideal energy level alignment for the most promising organic photovoltaics donor materials. Thus, this unprecedented screening study for organic photovoltaics not only identified the top candidates but also demonstrated design rules that are required to discover next-generation of high-performance photovoltaic materials. Furthermore, the techniques built in this work has laid a foundation for the active design and engineering of new molecular materials using HTPS for various applications.

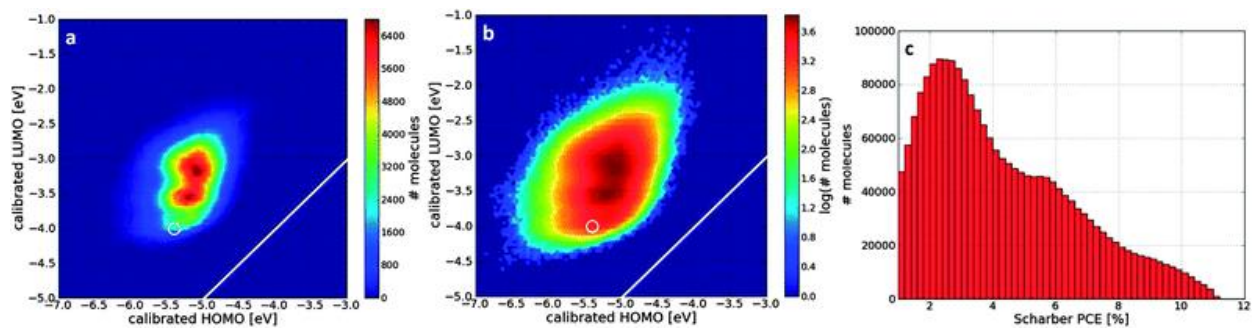

Figure 11: Screening results from the CEP project. HOMO and LUMO mapping of the 2.3 million molecular motifs on a linear (a) and logarithmic (b) scale. Panel (c) shows the resulting power conversion efficiency (PCE) histogram according to the Scharber model with respect to a phenyl-C61-butyric acid methyl ester acceptor. Figure reproduced from Ref. [15].

Another key challenge in the development of organic photovoltaics is the design of a suitable acceptor material [110, 115]. Fullerenes are the most commonly found acceptors in high-efficiency solar cells. However, very few donor materials energetically match with fullerenes, thus limiting the search space of donor molecules. One remedy is to develop alternative acceptor molecules which allow us to tailor the maximum LUMO energy offset. One of the alternatives is Acenes, widely studied for organic semiconductors, whose LUMO can be easily tailored by the addition of electron withdrawing substituents such as nitro, cyano and halide groups, or by substituting CH with nitrogen into the molecular framework [116, 117]. Halls *et al.* applied the latter approach to investigate the effect of nitrogen substituted pentacenes for potential electron acceptor materials [110]. The candidate library was developed using the MS Combi structure enumeration module [118] of Schrödinger Materials Science Suite [119]. The electronic properties were evaluated in an automated fashion using the Jaguar density functional theory (DFT) package [120]. The results from their screening study yielded few exemplary pentacene acceptor candidates, which were further evaluated with respect to their electron reorganization energies. The automated workflows, to explore the molecular design space, implemented in their work demonstrates its potential to efficiently accelerate the materials discovery process. Schrödinger's structure enumeration tool as well as the automated Jaguar workflow are extremely powerful, which has also shown to be highly promising for the discovery of materials in other applications [110, 121-126].

The organic-inorganic hybrid perovskites are another class of materials that have emerged as promising next-generation solar cells [127]. A recent spike in the interest of perovskite materials is due to their rapidly increasing PCE (4% PCE to 22% PCE in less than a decade) [128]. This performance enhancement is due to the intrinsic properties of halide perovskites whose monovalent cations are contained inside the cuboctahedral cell of metal halides ($MX_6$). Due to this structural symmetry and the direct band gap p-p transitions, halide perovskites exhibit excellent charge transfer properties, large absorption coefficient, and long electron-hole diffusion lengths. But, their short durability along with toxicity concerns is limiting their marketability. Fortunately, the crystal configuration of perovskites can be controlled to create new structures that are more durable and non-toxic along with enhanced optoelectronic properties. Depending on the cationic valence states and volume ratios the number of possible combinations is in the order of tens of thousands. For example, the number of combinations for (i) $AMX_3$ is greater than 24,000, (ii) $A_3M_2X_9$ is greater than 31,000, (iii) $A_2MX_6$ is greater than 22,000, and (iv) $A_2MM'X_6$ is greater than 9,000, where A is organic/inorganic component, M is metal, and X is a halogen element [107]. Thus, with a multitude of possible configurations, perovskites are an attractive material space for HTPS. Consequently, several HTPS efforts have been undertaken which include the screening of perovskites candidates in order of hundreds [111, 129-133]. A typical down-select recipe for the HTPS of perovskites includes the selection of perovskite structure (~$10^5$ structures), creating a library of possible configurations (~$10^4$ structures), stability screening (~$10^3$ structures), optical response calculations (~$10^2$ structures), charge mobility calculations (<100 structures), and excited state calculations (<40 structures), followed by experimental testing (<10 structures). In a very recent study, Nakajima *et al.* screened 11,025 configurations of $AMX_3$ and $A_2MM'X_6$ type perovskites [106]. Based on their screening studies, they identified 51 most promising halide single and double perovskites that are also environmentally friendly.

Another recent HTPS study by Korbel *et al.* [134] involved screening of more than 32,000 cubic perovskite combinations ($ABX_3$ type) to find thermodynamically stable compounds. They filtered the compounds down to 199 based on their photovoltaic, piezoelectric and magnetic properties, of which 128 were contained in current experimental databases and 71 were entirely new combinations with very high performance potential. This screening study further confirmed the promise of perovskite materials for solar applications.

Hautier *et al.* studied transparent conducting oxides (ternary and binary) to establish design rules on a much broader scale incorporating the influence of chemistry on overall conductivity, considering several atomic descriptors [135-137]. One of the major descriptors is effective mass, which is a cautious approximation for describing the mobility as noted by the authors themselves [138]. Furthermore, the localizing nature of electrons and holes in transition metal oxides is computationally inhibitive for HTPS study. A major area of research currently accounts for charge localization in such transition metal oxides instead of applying band transport approximation, so that screening studies generate efficient compounds.

Despite these large-scale HTPS efforts, the screening strategies for the discovery of new solar cell materials are still in their infancy. Nevertheless, these studies have established efficient protocols to screen for promising candidates and therefore can be extended to other perovskites as well as other class of materials [139-141]. Future HTPS studies can incorporate other properties (in addition to band gap and stability), such as band alignments, dielectric constants, and optical properties, in the rational design scheme to discover new-generation high-performance perovskites.

## b. Gas separation materials

As a result of burning fuels, there is a massive amount of $CO_2$ being released into the atmosphere. As it is difficult to cut the emissions, an easy and efficient approach to reduce $CO_2$ release is to separate and capture it from the mixture of gases from power plants like the flue gas. Several techniques including adsorption, absorption, distillation, and membrane-based separations have been used for gas separation. Recently, membrane-based technique attracted attention as it offers high efficiency, easy scale-up and is environmentally friendly. Metal-organic frameworks (MOFs)—crystalline porous materials composed of metal atoms and organic linkers—have interesting physical and chemical properties such as high porosities, large variation in pore size, and large surface area, which make them greatly promising for gas separation. MOFs with different pore sizes and shapes can be synthesized by controlling the combination of metal clusters and the organic blocks. Thus, HTPS approach is a systematic way to discover MOFs with high efficiency in $CO_2$ separation. Some of the HTPS studies undertaken in the past for the separation of $CO_2$ from other gas mixtures are listed in Table. 1. In addition to $CO_2$ separation, HTPS techniques are also being adapted to design materials for the separation of other gases such as $H_2$, $CH_4$, $NH_3$, and $I_2$.

Numerous computational models have been developed to evaluate gas separation/capture by quantifying Most commonly used metric to evaluate adsorbents is adsorption selectivity, the ratio of the content of adsorbed gas in adsorbent normalized by the ratio of bulk phase content of components [142, 143]. Other metrics that are also used for adsorption quantification are working capacity [144], adsorbent performance score [145], sorbent selection parameter [146], and regenerability. Working capacity, a typical metric for strong adsorbents, is the difference between the loading at the adsorption and desorption pressures. The adsorbent performance score is simply the product of the adsorption selectivity and working capacity. The sorbent selection parameter consists of two parts: ratio of the working capacity and the ratio of selectivity of the sorbent (for the strongly adsorbed species) at the equilibrium. Regenerability metric identifies the adsorption site regeneration when the desorption step is ongoing. These metrics have been used to rank the materials for gas adsorption. As these metrics are well defined, they have been implemented to select lead candidates in HTPS studies. Grand canonical Monte Carlo (GCMC), a combination of Monte Carlo and molecular dynamics method, is a typical computational approach to study the adsorption, diffusion, and permeation of gases into bulk materials. All the above defined metrics can be calculated using this method.

One of the first HTPS efforts of MOFs was performed by Wilmer and Snurr, in which they screened 130,000 MOFs for $CO_2$ separation [143]. This work established evaluation criteria for gas separation, while also presenting structure-property relations to guide the experimental synthesis of promising MOF candidates. In another study, 4,764 MOFs were screened for membrane separation of a $CO_2/N_2/CH_4$ gas mixture using the GCMC method [147]. Their work introduced a decision tree modeling technique to guide the screening, while principal component analysis and multiple linear regression methods were implemented to identify structure-property relationships. They identified seven MOFs which demonstrated efficient separation of both $CO_2$ and $N_2$ from $CH_4$, and therefore, are the most promising materials for upgrading natural gas. Another recent study applied GCMC to screen 3,857 MOFs to identify top candidates for the separation of $CO_2/H_2$ mixtures [148]. The structure-property relations from their study demonstrated that the best separation of $CO_2$ occurs for MOFs with low porosities and narrow pore sizes and the $H_2$ separation is dominant in membranes with high porosities and large pores.

A recent work by Altintas and co-workers included HTPS studies on MOFs for the separation of methane and hydrogen gas mixture [142]. They performed GCMC calculations on 4,350 MOFs, which were extracted from the Cambridge Structural Database, and ranked them for their performance on $CH_4/H_2$ separation. Several MOFs were shown to exhibit very high $CH_4/H_2$ selectivities in comparison to the conventional adsorbents including zeolites and activated carbons. In addition to identifying the top candidates, they also identified the structure-property relationships such as the relations between pore sizes, surface areas, heat of adsorption, adsorbility, and type of metal in MOF and their selectivities.

Table 1: HTPS efforts in the past to discover MOFs and zeolites for gas separation.

| Number of structures screened | Gas separation | Ref. |
| --- | --- | --- |
| 130,000 MOFs | $CO_2/CH_4$ | [143] |
| >10,000 zeolites | $CO_2/CH_4$ | [149] |
| 199 zeolites | $CO_2/CH_4$ | [150] |
| ~500 MOFs | $CO_2/N_2$ | [151] |
| ~300,000 zeolites | $CO_2/N_2$ | [152] |
| 1,800 zeolites library, 225 screened | $CO_2/N_2$ | [153] |
| 3,806 MOFs | $CO_2/N_2/H_2O$ | [154] |
| 4,764 MOFs | $CO_2/N_2/CH_4$ | [147] |
| 137,953 MOFs library, 17,257 screened | $CO_2/N_2/CH_4$ | [155] |
| 3,816 MOFs | $CO_2/N_2/CH_4$ | [156] |
| 5,109 MOFs library, 531 screened | $CO_2/H_2$ | [157] |
| 3,857 MOFs | $CO_2/H_2$ | [148] |
| 4,350 MOFs | $CH_4/H_2$ | [142] |
| 137,953 MOFs library, 2,777 screened | $NH_3$ capture | [158] |

Zeolites have also been extensively studied for gas separations, and several HTPS studies are reported in the literature, especially for $CO_2$ separation (see Table 1). In addition to MOFs and zeolites, Lin *et al.* performed HTPS to quantify the selectivity and capacity of $CO_2$ absorption in ionic liquids [159]. They used Henry's law constant, the product of infinite dilution activity coefficient of $CO_2$ molecule in the ionic liquid and the gas fugacity, as the ranking metric for the screening. They screened 2,080 ionic liquids, that were made by the combination of 65 cations and 32 anions, using this simple method. Their model is shown to be highly powerful and less computationally demanding, which could be used to study new ionic liquids for efficient $CO_2$ capture.

## c. Optical materials

Organic small molecules, oligomers, and polymers are emerging materials that feature many attractive properties in comparison to the conventional inorganic materials. Optical devices made from organic polymers are generally flexible, mechanically stable on impact, light-weight, and inexpensive to produce. This has resulted in increased efforts to utilize these compounds in many different application domains, including optic and optoelectronic devices in which they can be introduced *in situ* as microlenses, waveguides, microresonators, interferometers, anti-reflective coatings, optical adhesives, and substrates. However, most of these devices require materials with a refractive index (RI) greater than 1.7, while typical carbon-based polymers only exhibit values in the range of 1.3-1.5. This provides an incentive to discover or design new high-refractive index polymers (HRIPs) for the aforementioned applications. Since the properties of organic polymers can be tailored by controlling their molecular structure, they are prime candidates for a rational design target.

One of the prominent examples in the context of polymers for optoelectronic applications is the work by Ramprasad and co-workers [160, 161]. Their work included a screening of 1,073 polymers, which were collected from other existing sources, using *first-principles* DFT computations. From the screening studies, they computed the dielectric constant, atomization energy, and the energy band gap of these polymers. The primary goal of these screening studies was to generate sufficient data from the DFT calculations to be used to develop machine learning models [162]. These models were then used to evaluate the optical properties of thousands of new polymers. The methodology applied in this work can be extended to a different class of materials, given there is enough data to the train the model and easily extractable fingerprints are formulated.

In recent years, polyimides (PIs) have been shown to have favorable electronic and mechanical properties that could form potential HRIP candidates. Despite showing inherently low RI values leading to a lack of present applicability, PIs have other attractive properties [163, 164]. PIs exhibit exceptional thermal stability, and ease of processability [165, 166]. These properties are complemented by their favorable mechanical stability, flexibility, flame resistance, radiation resistance and their sufficiently high molecular polarizability: properties which would allow for potential use in optoelectronics [167, 168]. The optical properties of PIs can be improved by several methods [169]. One such technique is to control the chemical structure of PIs to allow for precise tuning of optical properties, in particular by increasing their RI values [170]. In a very recent study, we applied computational approach to study the RI of PIs and explore techniques that introduce highly polarizable moieties into polyimides framework to create a new class of high RI PIs. To facilitate RI evaluation of our large pool of candidates in a timely manner we used our virtual high-throughput screening framework, *ChemHTPS* [171]. *ChemHTPS* creates inputs, executes and monitors the calculations, parses and assesses the results, extracts and post-processes the information of interest, inserts the key outcomes into the project database, and archives all other data. This *in silico* methodology was implemented to create and characterize large number of PI candidates at a fraction of the time and cost of traditional studies.

The RI prediction model, used to characterize PI candidates, was based on a synergistic combination of *first-principles* calculations and machine learning [172, 173]. The model was validated using experimental RI values of 112 polymers, which shows that it is in an excellent agreement ($R^2$=0.94) with the experimental results. Figure 12a elucidates the relationship between polarizability and number density of these 112 polymers. The contours (for constant RI values) in this plot demonstrate an inverse proportionality relationship between the number density and the polarizability. Preferable high RI region happens when both the polarizability and number density are sufficiently high, i.e., towards the red contour line as shown in the Figure 12a. In our approach, we selected the structure of PIs such that the density is fairly constant, and increased the polarizability values by including highly polarizable aromatic structures.

Using the building blocks shown in Figure 12b, we created a library of $R_1$ and $R_2$ candidates and evaluated the RI values of these candidates by casting into *ChemHTPS*. We picked the top candidates of $R_1$ and $R_2$ and created a library of 100,000 PI candidates [171, 173, 174]. Most of the PI candidates exhibited RI values between 1.5 and 1.7, which means that there is a strong possibility of obtaining molecules with such RI values using empirical approaches. However, this screening study demonstrated that we can use computational techniques to identify candidates that possess RI values greater than 1.8. Other than identifying HRIP candidates, understanding the underlying structure-property relationships would enable us to discern candidates with optimal RI values. This would help us create a special subset of candidates for the experimental testing. To that end, we evaluated the contribution of each building block towards a targeted property to identify favorable building block. The results indicated that certain building block combinations are highly promising in the design of HRIPs. These design guidelines allowed us to target specific molecular motifs and create next generation polymers with exceptional optical properties.

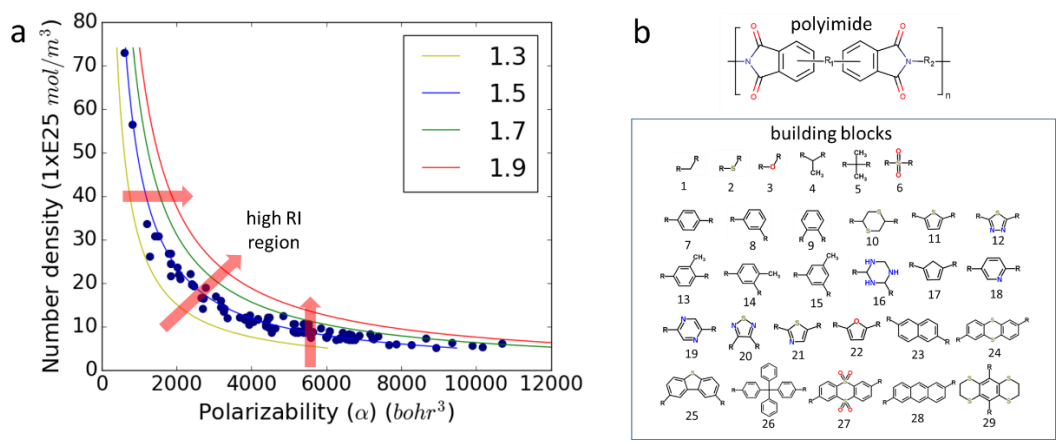

**Figure 12: a) Relationship between polarizability and number density along with the projection of 112 polymer values; b) Building blocks used to create the library of PIs. Figures adapted from Refs. [171, 172].**

### d. OLEDs

Organic light-emitting diode (OLED) technology is becoming highly attractive for lighting devices and display devices due to their superior color properties and efficiency. The light from OLEDs is emitted due to the relaxation of singlet excitons in the electroluminescent molecules that are present in the emissive layers. Current devices based on OLED technology use high-cost iridium-based metal complexes in these emissive layers. This is because iridium induces high spin-orbit coupling, which is vital for capturing all the singlet and triplet excitons, allowing for 100% energy transformation to light. However, there are two critical issues with the iridium complexes: the high cost of iridium and stability issue of these complexes for blue light emission, which limits their mass production. Therefore, an alternative OLED mechanism, thermally activated delayed fluorescence (TADF) [175], was proposed that allows for harvesting all singlet and triplet excitons in the lowest emitting singlet state [176]. The primary advantage of this mechanism is that the high efficiency can be achieved by the use of purely organic molecules, thus, making such devices significantly low-cost compared to Ir-based OLEDs.

For an organic molecule to be used as a successful TADF material, it should meet certain criteria [177]. These criteria include small separation between the triplet state and the singlet state, maintain a weak HOMO-LUMO overlap, homogenous HOMO-LUMO overlap over the TADF molecules, and fast TADF decay, which require both fast down-intersystem crossing and fast up-intersystem crossing. The above properties for potential organic molecules can be evaluated using DFT calculations, thus, allowing for computational screening of candidates. Furthermore, the ability to tailor the structure of organic molecules makes this highly attractive for HTPS. Consequently, several large-scale HTPS studies were undertaken in the past few years to identify promising organic candidates for TADF technology [126, 178-184].

One of the first HTPS efforts of organic molecules for OLEDs consisted of oligothiophene derivatives capped with different end groups [184]. The authors of this work used a semi-empirical method, PM6, to evaluate the HOMO, LUMO, and the band-gap of the library of molecules. Semi-empirical methods are less computationally demanding, thus allowing large-scale screening. Although the accuracy of these methods is poor, they can provide quick insights into the structure-property relationships and aid in reducing the chemical search space. This initial screening study suggest that the end-capped groups can be tailored to obtain targeted properties.

The molecules with TADF character consist of a donor block and an acceptor block, which are required for efficient thermal reverse intersystem crossing, i.e., for efficient thermal repopulation of emissive singlet state. The high efficiency is due to the low difference in the singlet-triplet gap caused by charge-transfer excitation. Thus, a simple recipe of donor-bridge-acceptor can be used to design new TADF molecules. This recipe was implemented by Aspuru-Guzik's group to create a library of 1.6 million organic molecules and performed one of the largest screening studies for OLEDs. As 1.6 million molecules are large for performing time-dependent density functional theory (TD-DFT) calculations, they use a machine learning approach to narrow the screening space. Initially, 40,000 randomly chosen molecules were screened and the resultant data was used to develop a neural network model. This model was subsequently applied to the complete library and the top candidates were subsequently validated using TD-DFT calculations. The new TD-DFT data was then included in the training set to develop new and efficient neural network models. Total 400,000 TD-DFT computations were performed in this screening studies. A single target property, upper bound on the delayed fluorescence rate constant, was used to rank the molecules. More than one thousand candidates were identified whose performance exceeds 22% external quantum efficiency. Such large-scale screening not only identified promising candidates but also provided chemical insight into the intrinsic limitations of TADF molecules.

Other areas where HTPS has been successfully applied include battery materials [185-189], photoabsorbers in water splitting [190, 191], scintillators and nuclear detection [192, 193], topological insulators [194, 195], piezoelectric materials [196, 197], viscoelastic materials [198], thermoelectric materials [199, 200], and magnetic materials [201].

## 4. Smart screening techniques

The down-select strategy, as shown in Figure 2, requires a large initial library of candidates. Instead of exploring a large chemical domain in its entirety, it is useful to narrow the search space to a region where candidates are most promising and synthetically viable. This can be achieved by augmenting the combinatorial schemes by a number of modules that make use of additional input. One approach is introducing constrained-growth schemes that continually prune the generation process to create more accessible or desirable candidates. In this scheme, molecules are rejected at every generation to limit the growth of molecules. The rejection could be based on constraints like the exclusion of certain structural patterns or substructures, fingerprint matching, building block combinations, or sequences. This approach can be seen in the several HTPS efforts that are reviewed in this chapter.

In many applications, due to enormous chemical space, applying above-mentioned generation constraints might still lead to a large number of unwanted molecules in the library. Applying smart algorithms, such as evolutionary algorithms, can narrow down the chemical space to a more targeted region. A schematic for the pruning of molecular libraries is shown in Figure 13. Genetic algorithm (GA) approach is one of the most commonly used algorithm for smart screening. GA employs an on-the-fly prescreening through rapid candidate assessment *via* DFT, MD, or data-derived prediction models. GA starts with an initial set of candidates and creates better candidates in every successive generation. Continuing the process for several generations results in a library that is tailored for the targeted application. Due to its efficiency in accelerating the process of materials discovery, GA approach gained significant attention in recent HTPS efforts [141, 202-210].

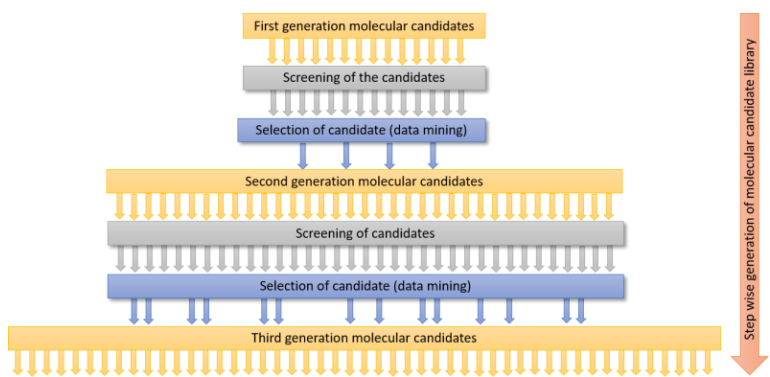

Figure 13: Schematic for the pruning of molecular library at every generation by use of smart algorithms.

In the context of finding ideal p- and n-type materials for organic photovoltaics, the challenge arises due to the enormous molecular space, estimated to be more than $10^{60}$ molecules. As it is not feasible to enumerate such a huge library, an efficient way is to reduce the space by applying GA. Hutchinson and co-workers applied GA to design and discover promising candidates for organic photovoltaics [205]. They further accelerated the process of screening by using fast computational methods in the initial steps of GA. Among the two million molecular combinations considered in their work, only 4% of the molecules were eventually sampled by GA. But, ~70% of those candidates were shown to have optimal properties, which shows a dramatic improvement over brute force methods.

In the case of MOFs, the key challenge lies in the selection of the optimal groups to functionalize for a particular application. The difficulty in the selection is due to the huge molecular space, e. g., the number of possible combinations could easily reach over two million when 40 functional groups are considered. Collins *et al.* applied GA approach to screen 1.65 trillion MOF structures for designing candidates with efficient $CO_2$ uptake [203]. The total number of structures that were finally selected by GA were about 500,000. Their study yielded more than a thousand structures that have exceptional $CO_2$ uptake (>3.0 mmol/g). In addition to the above two examples, smart algorithms have been successfully applied in the screening of materials for other applications, e.g., catalysis [207-209], conductors [211], and water splitting [141].

## 5. Conclusion and outlook

While conventional materials modeling and simulations are by now well-established elements in the toolbox of the materials research community, we can still consider virtual HTPS approaches an emerging technique. The former offer an efficient means to characterize material candidates and uncover promising targets for the more time-, labor-, and resource-intensive work in the laboratory. They can further provide a fundamental understanding of new findings that is outside the purview of empirical studies. Their focus on individual material candidate has, however, been limiting the utility of computational research. With the pursuit of virtual HTPS approaches, the community seeks to overcome some of these limitations.

The HTPS studies reviewed in this chapter demonstrate how new materials and catalysts with tailored property combinations can be discovered by this data-driven approach and how these new materials can facilitate a range of target applications. They also yield novel insights and systematic guidance to support the study and selection of highly promising domains in materials space.

The next frontier is to further develop the field of machine learning and data mining that is ideally suited to harness the large-scale data sets that result from computational HTPS studies. Despite some impressive pioneering work in this very young field, there is still a distinct disconnect between the promise of using modern data science in materials research and the realities of every-day research in the community, where data-driven work does not yet play a significant role. Advances in both HTPS and machine learning and the tight integration of these fields promises to have a considerable impact for the future of materials research.